\documentclass[aps,prp,twocolumn,superscriptaddress,raggedbottom,showpacs,floatfix]{revtex4}

\usepackage{amsmath,amsfonts,amssymb,amsthm,graphics}
\usepackage[next]{inputenc}
\usepackage[dvips]{epsfig}
\usepackage[colorlinks=true,citecolor=blue,linkcolor=blue]{hyperref}
\usepackage{bbm}
\usepackage{bbold}
\usepackage{booktabs}
\usepackage{multirow}
\usepackage{hhline}
\usepackage{amssymb}
\usepackage{comment}

\usepackage{bm}
\usepackage{color}
\usepackage{graphicx,latexsym}
\usepackage{epstopdf}

\begin{document}
\renewcommand{\vec}{\mathbf}
\renewcommand{\Re}{\mathop{\mathrm{Re}}\nolimits}
\renewcommand{\Im}{\mathop{\mathrm{Im}}\nolimits}
\renewcommand{\phi}{\varphi}

\title{Transport of Dirac electrons in a random magnetic field in topological heterostructures}
\author{Hilary M. Hurst}
\affiliation{Joint Quantum Institute and Condensed Matter Theory Center, Department of Physics, University of Maryland, College Park, Maryland 20742-4111, USA}
\author{Dmitry K. Efimkin}
\affiliation{Joint Quantum Institute and Condensed Matter Theory Center, Department of Physics, University of Maryland, College Park, Maryland 20742-4111, USA}
\author{Victor Galitski}
\affiliation{Joint Quantum Institute and Condensed Matter Theory Center, Department of Physics, University of Maryland, College Park, Maryland 20742-4111, USA}
\affiliation{School of Physics, Monash University, Melbourne, Victoria 3800, Australia}

\begin{abstract}
We consider the proximity effect between Dirac states at the surface of a topological insulator and a ferromagnet with easy plane anisotropy, which is described by the \emph{XY}-model and undergoes a Berezinskii-Kosterlitz-Thouless (BKT) phase transition. The surface states of the topological insulator interacting with classical magnetic fluctuations of the ferromagnet can be mapped onto the problem of Dirac fermions in a random magnetic field. However, this analogy is only partial in the presence of electron-hole asymmetry or warping of the Dirac dispersion, which results in screening of magnetic fluctuations. Scattering at magnetic fluctuations influences the behavior of the surface resistivity as a function of temperature. Near the BKT phase transition temperature we find that the resistivity of surface states scales linearly with temperature and has a clear maximum which becomes more pronounced as the Fermi energy decreases. Additionally at low temperatures we find linear resistivity, usually associated with non-Fermi liquid behavior, however here it appears entirely within the Fermi liquid picture.
\end{abstract}
\maketitle

\date{\today}

\maketitle
\section{Introduction}

The discovery of topological insulators (TI) has led to new ways to observe exotic physics in condensed matter systems, including phenomena such as magnetic monopoles and axion electrodynamics~\cite{Li2010,Qi2009}. Many of these insights rely on the nature of the TI surface states, which are described by the Dirac equation for relativistic particles (see \cite{Hasan2010, Qi2011} and references therein). The combination of TI and magnetic materials creates a hybrid platform to observe new physics by exploiting the spin-momentum locking of surface states. These dual-layer structures provide a way to experimentally realize disordered Dirac hamiltonians and localization phenomena in TI systems~\cite{Evers2008,Liao2015}.

Uniform out-of-plane magnetization, which can be induced by proximity effect or by ordered impurities deposited at the surface of a TI, opens a gap in the surface state spectrum~\cite{Chen2010}. This gapped state exhibits the anomalous quantum Hall effect, which can be  directly probed in transport or by magnetooptical Faraday and Kerr effects~\cite{Yu2010,Chang2013,Checkelsky2014,Tse2010,Maciejko2010,Efimkin2013,Wilson2014}. Out-of-plane magnetic textures such as domain walls and skyrmions host gapless chiral modes or localized states, altering their dynamics~\cite{Nomura2010,Ferreiros2014,Linder2014,Hurst2015}. As a result, the strong interplay of magnetism and surface states can be employed in spintronics applications.

In contrast, non-uniform in-plane magnetization can act as an effective gauge field. Dirac fermions exposed to transverse gauge field disorder, or a random magnetic field (RMF), have been studied theoretically in the context of the integer quantum Hall transition~\cite{Ludwig1994}, superconductivity~\cite{Minnhagen1987,Gusynin1999,Gusynin2002}, spin liquids~\cite{Galitski2005}, and disordered graphene~\cite{Khveshchenko2003} (see \cite{Evers2008} for a review). RMF disorder can strongly renormalize the spectrum and influence transport properties in both Schr\"{o}dinger and Dirac electron systems and leads to localization in the former~\cite{Altshuler1992,Aronov1994,Aronov1995,Taras2001}. A single Dirac cone will not localize in a short-range RMF, however whether localization occurs the case of long-range RMF presently lacks a definitive answer~\cite{Khveshchenko2007,Ostrovsky2007,Khveshchenko2008,Fedorenko2012,Konig2012,Fu2012,Essin2015}. The search for new experimental systems where the strength and spatial correlation of the magnetic field can be tuned is essential in the effort to understand the RMF problem. 

In this work, we consider such a system via the proximity effect between the surface of a TI and a thin-film magnet with easy plane anisotropy, which we describe by the \emph{XY}-model. The \emph{XY}-model enables magnetic vortex excitations and undergoes a Berezinskii-Kosterlitz-Thouless (BKT) phase transition, corresponding to vortex unbinding. We have shown that classical magnetic fluctuations of the \emph{XY}-model can be represented as an emergent static RMF acting on Dirac fermions, where the range of disorder is temperature dependent. Quasi-long-range gauge disorder below the BKT transition temperature is in general unscreened and can strongly influence Dirac states, making the problem intractable by usual perturbation methods.  However, we note that this gauge field analogy is not full in the presence of electron-hole asymmetry or warping terms, which depend on the doping level~\cite{Fu2009}. These terms lead to screening and the system can be tuned from the perturbative to nonperturbative regime by doping. We analyze transport in the doped, perturbative regime and we show that the resistivity has a prominent maximum near the BKT transition temperature, where magnetic fluctuations are the most intense. 

In Section II we present the model and discuss the mapping of classical magnetic fluctuations to static RMF disorder. In Section III we discuss the range of applicability of the perturbative treatment of the disorder. In section IV we calculate the temperature behavior of resistivity in the perturbative regime. In Section V we conclude and relate our work to current experiments. 
\section{Model}
\indent The purpose of this work is to show how signatures of an effective gauge field can be observed in transport of Dirac fermions. Coupling the Dirac states to a magnetic system with in-plane magnetic moments that undergoes a phase transition, like the 2D \emph{XY} model, allows for a system with tunable gauge disorder. The Dirac surface states of a 3D TI coupled to the \emph{XY}-model can be described by the following action 
\begin{equation}
S=S_{\mathrm{XY}}+S_{\mathrm{TI}}+S_\mathrm{TI}^\mathrm{XY}, 
\label{ModelGeneral}
\end{equation}
where 
\begin{align}
&S_{\mathrm{XY}}=\frac{\rho_\mathrm{s}}{2 T}  \int d\vec{r}~(\nabla \theta)^2; \quad  S_\mathrm{TI}^\mathrm{XY}=\Delta\int_0^\beta d\tau d\vec{r} ~\psi^\dagger \vec{n}(\vec{r})\cdot\boldsymbol{\sigma} \psi\nonumber;
\\
&S_\mathrm{TI}=\int_0^\beta d\tau  d\vec{r}~\psi^\dagger\left\{\partial_\tau +v\left[\vec{p}\times\boldsymbol{\sigma}\right]_z +\alpha p^2 - \mu \right\}\psi.
\label{Model}
\end{align}
The surface states are represented by a two-component spinor $\psi = (\psi_{\uparrow},\psi_{\downarrow})^T$, $\boldsymbol{\sigma} = (\sigma^x,\sigma^y)$ is the vector of Pauli matrices representing the real electron spin, $v$ is the Fermi velocity, and $\Delta > 0$ is the interlayer coupling between surface states and a magnetic \emph{XY}-model with magnetic moments $\vec{n}(\vec{r}) = \left(\cos(\theta),\sin(\theta)\right)$ with $\theta(\vec{r})$ describing their direction.

If electron-hole asymmetry is neglected ($\alpha = 0$), the magnetization plays the role of an emergent gauge field $\vec{a} = \Delta v^{-1}~[\vec{n}\times\hat{z}]$. It can be split as $\vec{a} = \vec{a}^\mathrm{l} + \vec{a}^\mathrm{t}$ into a transverse part, responsible for the emergent magnetic field $B_z=[\nabla \times \vec{a}^\mathrm{t}]_z=\Delta v^{-1} (\nabla \vec{n}^\mathrm{l})$ perpendicular to the surface, and a longitudinal part, which can generate an emergent electric field $\vec{E}=-\partial_t  \vec{a}^\mathrm{l}=-\Delta v^{-1}\partial_t \vec{n}^\mathrm{t}$. Here $\vec{n}^\mathrm{l}$ and $\vec{n}^\mathrm{t}$ are the corresponding components of spin density. Magnetic fluctuations are assumed to be classical, leading to zero emergent electric field, and therefore the longitudinal gauge field can be safely gauged away. 

The \emph{XY} model in 2D describes magnetic moments with fixed magnitude and arbitrary angle in the x-y plane. Low energy modes are described by the continuum model $S_\mathrm{XY}$ in equation (\ref{Model}) with temperature $T$ and spin-wave stiffness $\rho_\mathrm{S}$. The Mermin-Wagner theorem forbids long-range ordering in 2D at all nonzero temperatures~\cite{Mermin1966, Mermin1967}. However, the spin-spin correlation function exhibits unusual behavior, decaying algebraically at low temperatures $\propto \vec{r}^{-\eta}$ where $\eta(T) = T/2\pi\rho_\mathrm{s}$ is the critical exponent, which takes values from $\eta(0) = 0$ to $\eta_\mathrm{BKT}(T_\mathrm{BKT}) = 1/4$. The correlation function is $\propto \exp(-r/\xi_+)$ for $T > T_\mathrm{BKT}$. Near the transition the correlation length $\xi_+(T)$ is given by $\xi_+(T) \approx a\exp(3 T_\mathrm{BKT}/2\sqrt{T-T_\mathrm{BKT}})$, where a is cutoff for the magnet of the order of the vortex core size, which in turn is similar to the lattice constant of the magnet. $\xi_+(T)$ is finite only above the BKT transition and diverges exponentially as $T\rightarrow T^+_\mathrm{BKT}$~\cite{Berezinskii1971, Kosterlitz1973}. The transition between these two regimes, referred to as the BKT transition, is driven by the unbinding of magnetic vortex-antivortex pairs and occurs at $T_\mathrm{BKT} = \pi\rho_\mathrm{s}/2$. The \emph{XY}-model can occur in magnetic thin films with strong in-plane anisotropy, and has been realized in several compounds including $\mathrm{K}_2 \mathrm{Cu}\mathrm{F}_4$, Rb$_2$CrCl$_4$, BaNi$_2$(VO$_4$)$_2$ and (CH$_3$NH$_3$)$_2$CuCl$_4$~\cite{Hirakawa1982,Hirakawa1982b, Waibel2015,Regnault1983,Demokritov1990,Cornelius1986,Sachs2013}. 

Excitations of the magnetic \emph{XY}-model $\theta(\vec{r})=\theta_\mathrm{sw}(\vec{r})+\theta_\mathrm{v}(\vec{r})$ are spin-waves $\theta_\mathrm{sw}(\vec{r})$, creating a smooth emergent magnetic field $B_z$, and vortices $\theta_\mathrm{v}(\vec{r})$, which generate a nonunifrom magnetic field in a very nonlocal way. For a set of vortices situated at $\vec{r}_i$, the distribution of phase is $\theta_\mathrm{v}(z)=\sum_i q_i \arg(z-z_i)$ with $z=x+\bm{i}y$ and $q_i=\pm$ for vortices and antivortices. The resulting magnetic field is given by $B_\mathrm{z}=\Delta v^{-1} [\cos(\theta) \partial_y \theta_\mathrm{sw}- \sin(\theta) \partial_x \theta_\mathrm{sw} + \sum_i q_i \cos(\theta_i)/|\vec{r}-\vec{r_i}|]$ with $\theta_i=\theta_\mathrm{sw}(\vec{r}) + \sum_{j\neq i} q_j \arg[z-z_j]$. It diverges in the vicinity of each vortex core and its magnitude depends nonlocally on the position of all other vortices and slowly decays away from the vortex cores. 

The reconstruction of the local electronic structure due to the nonuniform spin density $\vec{n}(\vec{r})$ near the vortex core can be probed by tunneling experiments, as considered in detail in the case of magnetic impurities~\cite{Okada2011,Lee2015}. Here, we are interested in transport of Dirac fermions due to scattering at magnetic fluctuations where the chemical potential lies far above the Dirac point. In this case, scattering is restricted to the conduction band, and the vortex contribution to the effective magnetic field leads to an effective RMF as the conduction electrons see many vortices. 

The interaction between Dirac fermions mediated by spin fluctuations is obtained by integrating out the spin fluctuations and expanding to the second order in $\Delta/\mu$. The first order term vanishes; the second order term $S_\mathrm{d}$, corresponding to a disordered static magnetic field,  reads 
\begin{equation}
S_{\mathrm{d}}= - \frac{\Delta^2}{2} \int d\tau_1d\tau_2  d\vec{r}_1d\vec{r}_2 W_1^\alpha\left\langle
n^{\mathrm{l}}_\alpha(\vec{r}_1)n^{\mathrm{l}}_\beta(\vec{r}_2)\right\rangle W_2^\beta,
\label{eqn:Sd}
\end{equation}
where $W_i^\alpha =  \psi^{\dagger}(\vec{r}_i,\tau_i)\sigma^\alpha\psi(\vec{r}_i,\tau_i)$, and $\langle \cdots\rangle$ denotes averaging over the free \emph{XY} action including spin waves and vortices. The longitudinal part of the spin-spin correlation function is the only relevant one, given by 
\begin{equation}
\left\langle n^{\mathrm{l}}_\alpha(\vec{r}_1)n^{\mathrm{l}}_\beta(\vec{r}_2)\right\rangle = \frac{1}{2}\left(\frac{|\vec{r}_1 - \vec{r}_2|}{2a}\right)^{-\eta}\exp\left(-\frac{r}{\xi_+}\right)\Lambda^{\alpha \beta}_{\vec{r}_1 - \vec{r}_2},
\label{eqn:corr}
\end{equation}
where the matrix $\Lambda^{\alpha\beta}_\vec{q} = q_\alpha q_\beta/q^2$ ensures that only longitudinal spin fluctuations are taken into account and $a$ is the aforementioned lattice cutoff. Interaction between Dirac fermions $V_0^{\alpha\beta}(\vec{q}) =-\Delta^2\langle n^\mathrm{l}_\alpha(\vec{q})n^\mathrm{l}_\beta(-\vec{q})\rangle = V_0(\vec{q}) \Lambda_{\alpha \beta}$ is connected with the gauge invariant correlator of the emergent magnetic field $V_0(\vec{q})=-v^2 \langle B_z (-\vec{q}) B_z (\vec{q}) \rangle/q^2$ and is given by 
\begin{equation}
V_0(\vec{q})= -\frac{\pi\eta\Delta^2 \xi_+^{2-\eta}a^\eta }{(q^2 \xi_+^2+1)^{1-\eta/2}}.
\label{eqn:sprop}
\end{equation} 
For $T < T_\mathrm{BKT}$, $\xi_+ \rightarrow \infty$ and the propagator is $V_0(\vec{q}) \propto 1/q^{2-\eta}$, which results in singular behavior as $\vec{q}\rightarrow 0$ and strong temperature dependence through $\eta$. In 2D this leads to an infrared divergence in the self-energy which cannot be treated in a controlled manner in the absence of screening~\cite{Garate2010}. Particularly, in the random phase approximation (RPA) the screened interaction is given by
\begin{equation}
V^{-1}(q)=V_0^{-1}-\Pi^\mathrm{l}(q).
\end{equation}
Here $\Pi^\mathrm{l}(q)$ is the longitudinal spin-spin response function in the static limit which is given by~\cite{Garate2010}
\begin{equation}
\Pi^\mathrm{l}= \frac{q}{8\pi v}\mathrm{Re}\left[\frac{2k_\mathrm{F} }{q}\sqrt{1-\left(\frac{2k_\mathrm{F} }{q}\right)^2}+\arcsin\left(\frac{2k_\mathrm{F}}{q}\right)-\frac{\pi}{2}\right].
\end{equation} 
It is zero for $q \le 2k_\mathrm{F} $, which signals the absence of screening. Physically, vanishing $\Pi^\mathrm{l}(0)$ implies the absence of uniform spin polarization in the TI in the presence of a uniform external spin density $\vec{n}$. Really,  $\vec{n}$ can be safely gauged away through a transformation which shifts the position of a Dirac point $\vec{q}_\mathrm{D}=\Delta v^{-1} [\vec{n} \times \vec{e}_z]$, and therefore does not lead to any response.  

Above we have neglected electron-hole asymmetry ($\alpha=0$) in the Hamiltonian, describing electrons at the surface of the topological insulator. In Bismuth based topological insulators it is not negligible, but usually does not change the physics qualitatively (see~\cite{Baum2012, Pal2012} for an exception). Here we point out that the presence of electron-hole asymmetry or warping is crucial since it breaks the connection to the emergent gauge field picture. As a result, the spin-spin response function at low momenta becomes finite and allows screening. Recalling that $\Pi^\mathrm{l}(0)$ is the response to the uniform spin density, corresponding to momentum shift of Dirac states $\vec{q}_\mathrm{D}=\Delta v^{-1} [\vec{n} \times \vec{e}_z]$,  the resulting average spin polarization of electrons $\vec{s}_\mathrm{D} = \langle \psi^\dagger \boldsymbol{\sigma}\psi \rangle $ is given by 
\begin{equation}
\vec{s}_\mathrm{D}=\sum_{\vec{p}} \left[\vec{e}_\mathrm{z}\times  \frac{\vec{p}-\vec{q}_\mathrm{D}}{2 |\vec{p}-\vec{q}_\mathrm{D}|} \right] n_\mathrm{F}(\alpha p^2 + v |\vec{p}-\vec{q}_\mathrm{D}|-\mu),
\end{equation}
where $n_\mathrm{F}(\epsilon_p)$ is the Fermi step-like distribution at zero temperature. To linear order in $\vec{n}$ and $\alpha$ we get $ \vec{s}_\mathrm{D}=-\bar{\alpha} \vec{n} \nu_\mathrm{F} \Delta/2$ with $\nu_\mathrm{F}=\mu/2\pi\hbar^2v^2$ the density of states of Dirac electrons at the Fermi level and $\bar{\alpha}=\alpha \mu/v^2 $ the dimensionless electron-hole asymmetry strength, which leads to $\Pi^\mathrm{l}(0) = -\bar{\alpha}\nu_\mathrm{F}\Delta/2$. The spin polarization vanishes in the absence of electron-hole asymmetry as expected. The screened interaction in RPA mediated by magnetic fluctuations at $T < T_\mathrm{BKT}$ is then given by
\begin{equation}
V(\vec{q}) = -\frac{\pi\eta \Delta^2\xi_\alpha^{2-\eta} a^{\eta}}{( q\xi_\alpha)^{2-\eta}+1}\mbox{~~;~~}\xi_{\alpha} =\left( \frac{2}{\bar{\alpha}\nu_\mathrm{F}\Delta\pi\eta a^\eta}\right)^{\frac{1}{2-\eta}}
\label{eqn:Vscreen}
\end{equation}
The strength of electron-hole asymmetry, which regularizes our theory, can be characterized by the dimensionless parameter $\bar{\alpha}$ which decreases with the chemical potential $\mu$ and vanishes in the undoped regime. As a result, by controlling the doping level the system can be tuned from the perturbative to non-perturbative regime. To clarify the range of applicability of the perturbative approach, we consider the renormalization of the single-particle spectrum. 

\section{Self-energy of Dirac electrons}
In the Born approximation, the self energy of Dirac electrons is given by
\begin{equation}
\Sigma^\mathrm{R}(\omega,p) = \int_{q} \mathcal{Q}^\alpha_{\vec{p},\vec{p}-\vec{q}} V^{\alpha\beta}(\vec{q})\mathcal{Q}^\beta_{\vec{p}-\vec{q},\vec{p}}G_0^\mathrm{R}(\omega,\vec{p}-\vec{q}),
\label{eqn:SE1}
\end{equation}
where $\bm{\mathcal{Q}}_{\vec{p},\vec{p}'} = \langle p|\bm{\sigma}|p'\rangle = ( -\sin[(\phi_\vec{p} + \phi_{\vec{p}'})/2],  ~\cos[(\phi_\vec{p} + \phi_{\vec{p}'})/2])^T$ is the matrix element for scattering of electrons from the conduction band and their Green's function is $G_0^\mathrm{R} = (\omega - v p +\mu+ \bm{i}\delta)^{-1}$. Summation over $\alpha,\beta = x,y$ in equation (\ref{eqn:SE1}) gives the angle factor $\bar{\mathcal{Q}}_{\vec{p},\vec{p}-\vec{q}} = \mathcal{Q}^\alpha_{\vec{p},\vec{p}-\vec{q}} \Lambda ^{\alpha\beta}_\vec{q}\mathcal{Q}^\beta_{\vec{p}-\vec{q},\vec{p}}$ as follows 
\begin{equation}
\bar{\mathcal{Q}}_{\vec{p},\vec{p}-\vec{q}} = \sin^2\left(\frac{2\phi_\vec{q}-\phi_\vec{p}-\phi_{\vec{p}-\vec{q}}}{2}\right),
\end{equation}
where $\phi_{p}$ denotes the polar vector of a fermion with momentum $\vec{p}$. If the scattering is elastic $|\vec{p}| = |\vec{p}-\vec{q}|$, trigonometry dictates $2\phi_\vec{q}-\phi_\vec{p}-\phi_{\vec{p}-\vec{q}}=\pi$ and $\bar{\mathcal{Q}}_{\vec{p},\vec{p}-\vec{q}}=1$. $\Re\Sigma(0,p_\mathrm{F})$ at the Fermi level leads to Fermi energy renormalization, and it is zero for this case. Inserting the screened propagator, equation (\ref{eqn:Vscreen}), into (\ref{eqn:SE1}) gives a single particle decay rate $\hbar\gamma= -\Im\hat{\Sigma}(0,p_\mathrm{F})$, where 
\begin{equation}
\Im\hat{\Sigma} = -\frac{\Delta^2\eta\xi_\alpha^{1-\eta}a^\eta}{2\pi\hbar v}~
\Gamma\left(\frac{3-\eta}{2-\eta}\right)\Gamma\left(\frac{\eta-1}{\eta-2}\right).
\label{eqn:SE2}
\end{equation}
The product of gamma functions is of order 1 for $0 < \eta < 1/4$.
\begin{figure}
\centering
\includegraphics[width = 0.5\textwidth]{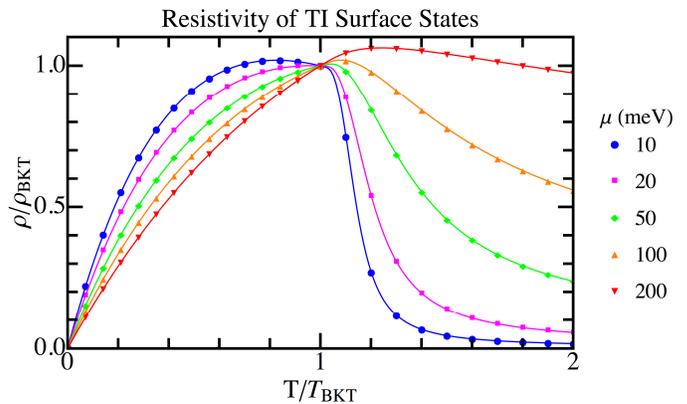}
\caption{(color online)~Resistivity of TI surface states coupled to an \emph{XY} model. The resistivity scales linearly with temperature as $T \rightarrow 0$ and across the BKT transition, with a nonuniversal peak at $T \sim T_\mathrm{BKT}$ that increases with increasing $\mu$. As $\mu$ increases the effect of the transition is less pronounced.\label{Fig:Resist}}
\end{figure}

The single particle life time diverges in the absence of screening, where $\xi_{\alpha}\rightarrow \infty$ for  $\bar{\alpha}\rightarrow 0$, as expected. It cannot be cured by using the self-consistent Born approximation and signals the break down of the perturbative approach, which we use below for a  calculation of conductivity of Dirac fermions. The apparent breakdown of perturbation theory in this model could also signal the existence of a non-Fermi liquid state on the TI surface, a possibility which could be explored by including the dynamics of the BKT magnet, which is beyond the scope of this work.

In the presence of screening, the Fermi-liquid approach breaks down for $\hbar\gamma \gtrsim \mu$. Using equations (\ref{eqn:Vscreen}) and (\ref{eqn:SE2}) we find a lower bound on the Fermi energy
\begin{equation}
\mu_c^{4-3\eta} \simeq \frac{\Delta^3 \eta v^2}{\pi^2\alpha}\left(\frac{2\Delta\hbar v^3}{\pi \alpha a}\right)^{-\eta}.
\end{equation}
For $\mu < \mu_c$ this approach is no longer valid and the system could be tuned from the perturbative to nonperturbative regime by doping. We estimate for $\mathrm{Bi}_2\mathrm{Te}_3$, $v = 0.5\times10^6~\hbox{m/s}$, $\Delta = 10~\hbox{meV}$ and $\alpha = 1/2m^*$ with $m^* \sim 0.1 m_e$. At $T=T_\mathrm{BKT}$, $\eta_\mathrm{BKT} = 1/4$ which gives $\mu_c\gtrsim 6~\hbox{meV}$. However, it is also important to keep in mind that for $\mu \lesssim \Delta$ higher order terms in the expansion (\ref{eqn:Sd}) become important and are not considered here. We leave the non-perturbative regime to further investigations which could be informed by this type of experiment. 
\section{Transport of Dirac fermions}
In the doped regime at $\mu\gg \hbar\gamma$, the quasipartcle picture is well defined and the resistivity of Dirac fermions can be approximated by the Drude formula
\begin{equation}
\rho=\frac{h}{e^2} \frac{2\hbar}{ \mu \tau_\mathrm{tr}},
\label{eqn:rho}
\end{equation}
where $\tau_\mathrm{tr}$ is the transport scattering time. Different scattering mechanisms, including impurities, phonons, and spin-fluctuations, additively contribute to $\tau_\mathrm{tr}^{-1}$ and can be easily separated. Here we concentrate on elastic scattering due to magnetic fluctuations, where for $|q| = 2k_\mathrm{F}\sin\phi/2$ the corresponding contribution is given by  \begin{equation*}
\frac{1}{\tau_\mathrm{tr}}= \frac{2\pi}{\hbar}\int_{q} \bar{\mathcal{Q}}_{\vec{p},\vec{p}-\vec{q}}\left|V_\vec{q}\right|\left(1-\cos\phi_{q}\right)\delta(\xi_{\vec{p}-\vec{q}}-\xi_\vec{p})=
\end{equation*}\vspace{-20 pt}
\begin{equation}
=\frac{\pi \eta \Delta^2}{4\hbar \mu} \left(\frac{\mu}{\mu_\mathrm{a}}\right)^\eta \int \frac{d\phi}{\pi}\frac{(2k_\mathrm{F}\xi_{+})^{2-\eta}\sin^2(\frac{\phi}{2})}{\left[(2 k_\mathrm{F} \xi_+\sin(\frac{\phi}{2}))^2 + 1\right]^{\frac{2-\eta}{2}}},
\label{eqn:tautr}
\end{equation}
with $\mu_\mathrm{a}=\hbar v/2a$. In contrast to the single particle decay rate $\gamma$, the inverse transport time $\tau_\mathrm{tr}^{-1}$ does not diverge in the absence of screening and weakly depends on screening length $\xi_\alpha$. Therefore, in the equation (\ref{eqn:tautr}) we used the unscreened propagator $V_0(\vec{q})$ given by equation (\ref{eqn:sprop}). Nevertheless, we need to keep in mind that  the derivation of Drude formula implies $\hbar\gamma\ll\mu$ since all diagrams with crossed impurity lines, which are important in the opposite regime, are neglected ~\cite{Bruus2004}.
Using equation (\ref{eqn:rho}) and the results above, the resistivity has the form
\begin{equation}
\frac{\rho(T,\mu)}{\rho_\mathrm{BKT}} = \frac{\eta}{I_0\eta_\mathrm{BKT}}\left(\frac{\mu}{\mu_a}\right)^{\eta-\eta_\mathrm{BKT}}\mathcal{I}(\eta,\phi).
\label{eqn:resist}
\end{equation}
Where $\mathcal{I}(\eta,\phi)$ is the integral in equation (\ref{eqn:tautr}), $\eta_\mathrm{BKT} = 1/4$, and $I_0 = \mathcal{I}(\eta_\mathrm{BKT},\phi) \approx 1.72$. The resistivity at the transition is given by 
\begin{equation}
\rho_\mathrm{BKT}= \frac{h}{e^2}\frac{\sqrt{\pi}\Delta^2}{4~\mu^2}\left(\frac{\mu}{\mu_a}\right)^{\frac{1}{4}}\frac{\Gamma(5/8)}{\Gamma(9/8)}.
\end{equation} 
$\rho(T,\mu)/\rho_\mathrm{BKT}$ is shown in Figure \ref{Fig:Resist} for different values of $\mu$. There is a clear peak near $T_\mathrm{BKT}$ due to increased magnetic fluctuations. As $T\rightarrow 0$ we find the following expression
\begin{equation}
\frac{\rho(T\rightarrow 0)}{\rho_\mathrm{BKT}} = \frac{\sqrt{\pi}~\Gamma(9/8)}{\Gamma(5/8)}\left(\frac{\mu}{\mu_a}\right)^{-\frac{1}{4}}\frac{T}{T_\mathrm{BKT}},
\end{equation}
where the resistivity is linear at low temperature, unlike usual impurity scattering. As $T\rightarrow T_{BKT}^{\pm}$ across the transition, we find that
\begin{align}
\frac{\rho(T\rightarrow T^+_\mathrm{BKT})}{\rho_\mathrm{BKT}} &= 1 + \frac{1}{4}\left\lbrace 4+\ln\left(\frac{\mu}{\mu_a}\right)\right\rbrace\frac{\Delta T}{T_\mathrm{BKT}},\\
\frac{\rho(T\rightarrow T^-_\mathrm{BKT})}{\rho_\mathrm{BKT}} &= 1 + \frac{1}{8}\left\lbrace 8 + 2\ln\left(\frac{\mu}{\mu_a}\right)\right. \\ 
& +\left.\psi\left(\frac{5}{8}\right)-\psi\left(\frac{9}{8}\right)\right\rbrace\frac{\Delta T}{T_\mathrm{BKT}}\nonumber,
\end{align}
where $\Delta T =T-T_\mathrm{BKT}$ and $\psi$ is the digamma function. The resistivity is linear in temperature in all three regimes but has a different slope in each case. As $T\rightarrow 0$, the dynamics of the magnetic moments becomes important, necessitating a fully quantum theory which is not considered here. The slope is dictated by $\mu$ and changes significantly with doping as shown in Figure \ref{Fig:Resist}. The temperature of maximal resistivity is also dictated by $\mu$; it occurs for $T \sim T_\mathrm{BKT}$ but is nonuniversal. It can be solved for using the exact expression in equation (\ref{eqn:resist}). We note that for $\mu < 13~\hbox{meV}$ the slope is always negative for $T > T_\mathrm{BKT}$ and the maximum resistivity occurs before the transition. 

In a real experiment there will be many sources of scattering including phonons and non-magnetic impurities. The linear temperature dependence and sharp peak near the BKT transition enables the separation of this scattering mechanism from others in the system. Scattering due to impurities is temperature independent, while at low temperatures phonon scattering leads to a different scaling law. 

\section{Conclusion and Discussion}
Physical realization of Dirac fermion-gauge field models in TI systems relies heavily on the strength of magnetic perturbations to the TI system. In this section we provide some estimations of the coupling strength $\Delta$ and how it connects to current experiments. For the 3D TI Bi$_2$Te$_3$ we use $\mu = 0.1~\hbox{eV}$. The transport lifetime in Bi$_2$Te$_3$ can be inferred from transport measurements to be $\tau_0 \sim 10^{-12}~\hbox{s}$~\cite{Qu2010}. The short distance cutoff $a$ is estimated by the half the lattice constant of two dimensional BKT magnet K$_2$CuF$_4$, where $a \sim 2.5~$\AA~\cite{Hirakawa1982b}.

In order to observe the anomalous transport behavior described above, the coupling between the magnetic layer and the TI must be strong enough such that $\tau_{tr} \lesssim \tau_0$. The transport time $\tau_{tr}$ at $T = T_\mathrm{BKT}$ is found from equation (\ref{eqn:tautr}), where  
\begin{equation}
\tau^{-1}_{tr} = \frac{\pi\Delta^2}{16\hbar\mu}\left(\frac{\mu}{\mu_a}\right)^{\frac{1}{4}}I_0.
\end{equation}
Setting $\tau^{tr} = \tau^0$ gives a lower bound on the coupling strength $\Delta$. For our parameters, we find $\Delta \gtrsim 10~\hbox{meV}$, which is well within the range of the recently observed $\Delta\sim 85~\hbox{meV}$ in lanthanide-doped Bi$_2$Te$_3$~\cite{Harrison2015}.

To summarize, we have considered the transport of Dirac fermions coupled to an \emph{XY}-model as temperature is tuned through the BKT transition. We claim that both short-range and quasi long-range disorder can be realized, and the transition between these regimes can be tuned by both doping level and temperature, thus determining the strength and nature of the disorder. We have analyzed the resistivity at high doping and we find that it scales linearly with temperature, with a prominent peak at the BKT transition temperature where magnetic fluctuations are the strongest. Notably, the resistivity also scales linearly with temperature as $T\rightarrow 0$. The effect is strengthened by decreasing the Fermi energy. 

\section{Acknowledgements}
The authors are grateful to Boris Altshuler for valuable discussions. This research was supported by DOE-BES DESC0001911 and the Simons Foundation. H.H. acknowledges additional fellowship support from the National Physical Science Consortium and NSA.
\bibliographystyle{apsrev}
\bibliography{main}

\end{document}